\documentclass[pre,
aps,
a4paper,
english,
showpacs,
reprint,
twocolumn,
superscriptaddress,
floatfix
]{revtex4-2}

\usepackage{multirow}
\usepackage{adjustbox}
\usepackage{makecell}
\usepackage{tabularx}
\usepackage{array}
\usepackage{graphicx}
\usepackage{mathtools}
\usepackage[dvipsnames]{xcolor}
\usepackage{tikz}
\usepackage{wrapfig}
\usepackage[T1]{fontenc}
\usepackage[utf8]{inputenc}
\usepackage{amsmath}
\usepackage{amssymb}
\usepackage{blkarray,booktabs,bigstrut}

\usepackage[
    citecolor=blue,
    colorlinks,
    linkcolor=blue,
    urlcolor=blue
]{hyperref}

\usepackage{bm} 
\usepackage{enumerate}
\begin{document}
	
	
\title{{Extending Quantum Perceptrons: Rydberg Devices, Multi-Class Classification, and Error Tolerance}}

\author{Ishita Agarwal}
\affiliation{Department of Computing + Mathematical Sciences (CMS), California Institute of Technology (Caltech), Pasadena, CA 91125, USA}
\affiliation{Department of Physics, Indian Institute of Technology Kanpur, Uttar Pradesh 208016, India}

\author{Taylor L. Patti}
\affiliation{NVIDIA, Santa Clara, CA 95051, USA}

\author{Rodrigo Araiza Bravo}
\affiliation{Department of Physics, Harvard University, Cambridge, Massachusetts 02138, USA}

\author{Susanne F. Yelin}
\affiliation{Department of Physics, Harvard University, Cambridge, Massachusetts 02138, USA}

\author{Anima Anandkumar}
\affiliation{Department of Computing + Mathematical Sciences (CMS), California Institute of Technology (Caltech), Pasadena, CA 91125, USA}

\date{\today}

\vspace{0.5cm}

\begin{abstract}
   Quantum Neuromorphic Computing (QNC) merges quantum computation with neural computation to create scalable, noise-resilient algorithms for quantum machine learning (QML). At the core of QNC is the quantum perceptron (QP), which leverages the analog dynamics of interacting qubits to enable universal quantum computation. Canonically, a QP features $N$ input qubits and one output qubit, and is used to determine whether an input state belongs to a specific class. Rydberg atoms, with their extended coherence times and scalable spatial configurations, provide an ideal platform for implementing QPs. In this work, we explore the implementation of QPs on Rydberg atom arrays, assessing their performance in tasks such as phase classification between Z2, Z3, Z4 and disordered phases, achieving high accuracy, including in the presence of noise. We also perform multi-class entanglement classification by extending the QP model to include multiple output qubits, achieving 95\% accuracy in distinguishing noisy, high-fidelity states based on separability. Additionally, we discuss the experimental realization of QPs on Rydberg platforms using both single-species and dual-species arrays, and examine the error bounds associated with approximating continuous functions.
\end{abstract}

\maketitle
 
\section{Introduction}

Quantum computing and neuromorphic computing are two of the most promising paradigms for computational innovations. Quantum computing seeks to harness quantum resources such as entanglement and superposition to develop algorithms that can outperform classical methods~\cite{steane1998quantum, horowitz2019quantum}. On the other hand, neuromorphic computing draws inspiration from the brain's neural networks to achieve efficient, low-energy computation~\cite{furber2016large, markovic2020physics}. Recently, the convergence of these two fields has given rise to quantum neuromorphic computing (QNC)~\cite{Markovic2020, Mujal2021}, a multidisciplinary approach that merges the principles of quantum mechanics with the dynamics of neural computation. QNC exploits the similarities between the behavior of neurons in the brain and the dynamics of noisy qubits, striving to deliver noise-resilient, scalable algorithms for quantum machine learning (QML)~\cite{Martinez-Pena2020, Bravo2022}. A central challenge in this subfield is the development of simple, scalable models for neuron-like dynamics, which would serve as building blocks for quantum neuromorphic systems. \par
The quantum perceptron (QP)~\cite{Freund1998} emerges as a foundational unit in QNC, utilizing the analog dynamics of interacting qubits with tunable couplings and single-qubit rotations. The model of Bravo et al.~\cite{AraizaBravo2023} has demonstrated significant potential in addressing various QML challenges, such as computing inner products of complex quantum states~\cite{buhrman2001quantum}, detecting quantum entanglement~\cite{dowling2004energy,toth2005entanglement,sperling2013multipartite}, and advancing quantum metrology~\cite{giovannetti2011advances}. It has also been established that QPs offer computational expressiveness by approximating any unitary operation and mitigating training challenges like barren plateaus through entanglement thinning \cite{AraizaBravo2023}. While several QP models have been proposed, many rely on currently infeasible architectures~\cite{Kapoor2016, Kerstin2019}, slow adiabatic processes~\cite{Torrontegui2019, Ban2021}, or challenging multiqubit topologies~\cite{Gao2022}. In contrast, the model of Bravo et al.~\cite{AraizaBravo2023} used in this pa\-per is based on rich, yet experimentally tractable, Hamiltonian interactions, making it well-suited for near-term quantum devices and practical QML applications.\par
\begin{figure*}[ht]
    \centering
\includegraphics[height=0.4\textheight]{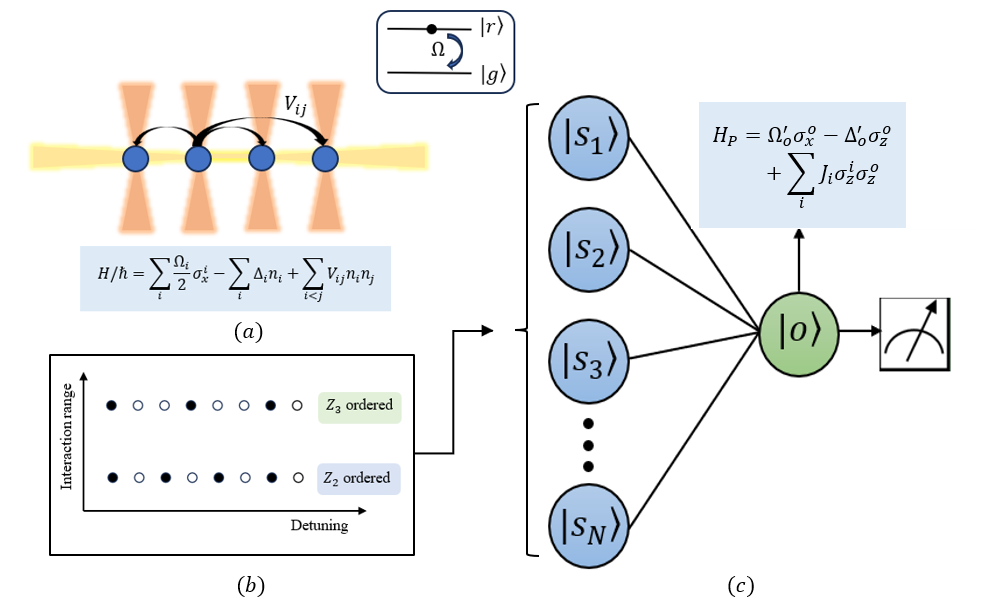}
    \caption{(a) Individual \(^{87}\text{Rb}\) atoms are trapped using optical tweezers (vertical red beams) and and arranged into defect-free arrays with probabilistic configuration. Coherent interactions \(V_{ij}\) between the atoms are facilitated by exciting them to a Rydberg state with interaction strength \(\Omega\) and detuning \(\Delta\). (b) The schematic illustrates the ground-state phase diagram of the Hamiltonian, highlighting phases \(Z_2\) and \(Z_3\) with different broken symmetries based on interaction range and detuning. A dataset of noisy states from these phases serves as the perceptron's input. (c) The QP comprises \( N \) input qubits and a single output qubit. The qubits undergo evolution governed by a Hamiltonian, ensuring that the probability of the output qubit being in the \( |0\rangle_0 \) state is a nonlinear function of the state of the input qubits. Following this evolution, the output qubit is measured \cite{51Atom2023}.}
    \label{fig:wide_figure}
\end{figure*}
One of the major challenges of controlling many-body quantum systems is the tendency of interactions to cause thermalization and chaos in the system's quantum states~\cite{deutsch1991quantum, srednicki1994chaos}. However, recent breakthroughs in the controlled manipulation of isolated, many-body quantum systems have permitted the probing of nonequilibrium states that are far beyond the reach of classical numerical simulations~\cite{Bluvstein2021,Ladd2010}. Arrays of neutral atoms have gained significant interest by incorporating Rydberg states, forming what is known as Rydberg atom arrays~\cite{51Atom2023}. These arrays are well-suited for studying complex quantum systems due to the strong dipole-dipole interactions, large polarizability and extended coherence times of Rydberg atoms, enabling fast and controllable operations across various geometrical configurations~\cite{Rydberg2023}.  While other potential experimental platforms for QPs include optical lattices~\cite{Masson2020, Patti2021, Castells-Graells2021}, and nitrogen-vacancy centers in diamonds~\cite{Bradley2019}, Rydberg atoms are especially promising candidates for implementing QPs because of their scalable spatial configurations and the high level of control they offer over individual qubits. In this manuscript, we explore the fundamental mechanics necessary for implementing a QP on Rydberg atom arrays and discuss strategies to encode them experimentally: using specific atomic arrangements to exploit Van der Waals potentials and flip-flop Hamiltonians in single-species arrays, and employing dual-species arrays which minimize cross-talk between input qubits.\par
In addition to mapping QPs to Rydberg atom arrays, we also explore key aspects and capabilities of QPs. We demonstrate their robustness as classifiers by successfully distinguishing between quantum phases and performing entanglement classification under varying noise conditions. To enable scalable and efficient multi-class classification, we extend the QP model to incorporate two output qubits, rather than one output qubit as previously studied, laying the foundation for their integration into deeper neural network architectures. Additionally, we quantify the approximation error bounds of QPs, using recent theoretical advancements to provide a lower-bound on the approximation error for parameterized quantum circuits when approximating continuous functions with integrable Fourier transforms.\par
The structure of the manuscript is as follows. In Section~\ref{sec2a}, we define QPs through their unitary dynamics and the tunable nonlinear activation functions that this architecture enables. In Section~\ref{sec2b}, we explore the underlying learning theory of QPs, emphasizing their potential in QML applications. Section~\ref{sec2c} focuses on the dynamics of Rydberg atoms, setting the stage for Section~\ref{sec:mapping}, where we demonstrate how our perceptron can be realized using arrays of Rydberg atoms by mapping their Hamiltonians. Section~\ref{exp} addresses the experimental realization of the QP within arrays of Rydberg atoms. In Section~\ref{sec_phase}, we evaluate the performance of QPs in phase classification tasks under varying noise levels. The extension of the QP model to include two output qubits is discussed in Section~\ref{sec2oq}, where we also assess its performance in multi-class classification. Finally, in Section~\ref{errorbound}, we discuss the approximation error bounds associated with QPs, and Section~\ref{conclusion} concludes the paper by summarizing our findings and discussing future directions.
\section{Preliminaries}
\subsection{Quantum Perceptrons}\label{sec2a}
\begin{figure*}[ht]
    \centering
    \includegraphics[width=\textwidth]{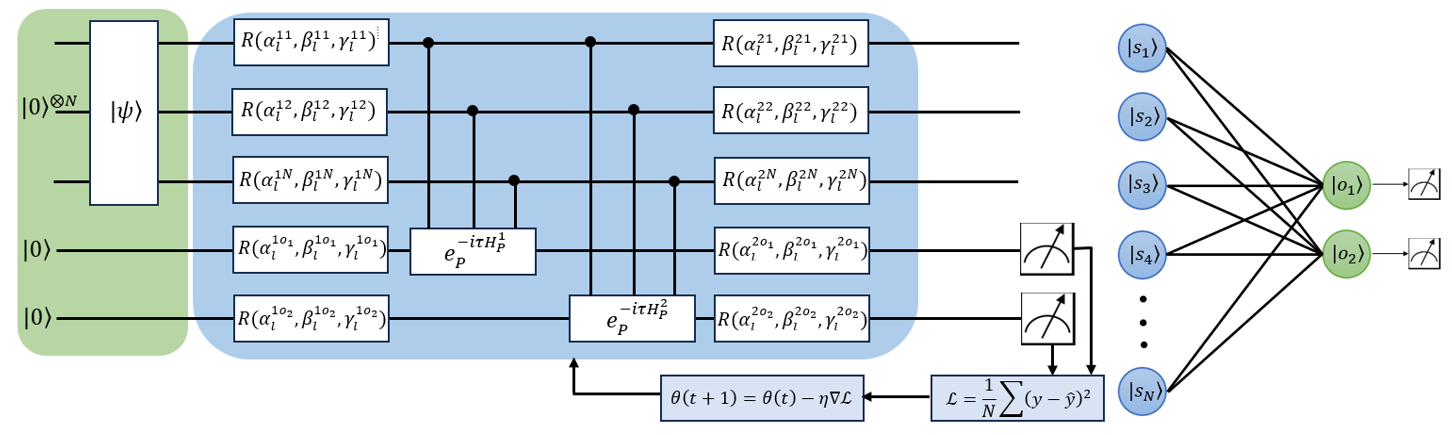}
    \caption{Schematic of a QP that operates on \(N\) input qubits and 2 output qubits (right) which evolve under the Hamiltonian in Eq. (\ref{2oq_perc}).  The circuit (left) begins with the preparation of the input state \(|\Phi(x)\rangle\) (green) where each input \(x\) is encoded into the quantum state. The input-output system then evolves via a series of single-qubit rotations along the x-, y- and z-axes, interspersed with controlled entangling gates, as indicated by the Hamiltonian \(H_P\) for a time \(\tau\). Each output qubit interacts independently with all input qubits, ensuring that their evolution is solely influenced by the input qubits without interacting with each other. Finally, the output of the circuit is measured to obtain the expectation values, which are then used to calculate a loss function. This loss function compares the obtained expectation values to the target function \(\tilde{y}(x)\) defined by the labels. The parameters of the QP are updated via gradient descent to minimize this loss, optimizing the performance of the perceptron for multi-class classification tasks. 
 }
    \label{fig:2oq_circuit}
\end{figure*}
A QP is a quantum-mechanical extension of a classical perceptron, a simplified biological neuron model used in machine learning and artificial neural networks~\cite{mcculloch1943logical} for classification tasks. 
In formal terms, a QP is defined as a set of $N+1$
qubits with $N$ input qubits and a single output qubit 
denoted as $o$. Each of the $N$ input qubits is analogous to the input neurons in a classical perceptron, and they interact with the output qubit $o$ through Hamiltonian mediated quantum dynamics. FIG.~\ref{fig:wide_figure} (c) shows a schematic of the QP. The entire quantum system resides in the product Hilbert space \( \mathcal{H} = \bigotimes_{n=1}^{N+1} \mathcal{H}_n \), where each individual qubit occupies its own Hilbert space $\mathcal{H}_n$ spanned by the basis vectors \( \left\{ \left| 0 \right\rangle_n, \left| 1 \right\rangle_n \right\} \). The dynamics of this quantum system are dictated by the Hamiltonian~\cite{AraizaBravo2023}:
\begin{equation}\label{perc_eq}
   H_P = -\Delta_o' \sigma^o_z + \Omega_o' \sigma^o_x + \sum_{i=1}^{N} J_i \sigma^i_z \sigma^o_z.
\end{equation}
where the parameter \(\Delta_o\) is the detuning of the output qubit with the driving field \(\Omega_o\), a coherent drive that couples the states \(|0\rangle_o\) and \(|1\rangle_o\) of the output qubit, \(\sigma^n_z=|0\rangle\langle0|_n - |1\rangle\langle1|_n\) is the Pauli-Z operator and \(\sigma^n_x=|0\rangle\langle1|_n + |1\rangle\langle0|_n\) is the Pauli-X operator. The parameters \(\{J_i\}_i\) denote the coupling strengths between the input and output qubits.

\subsection{Learning with Quantum Perceptrons}\label{sec2b}

Equipping the QP with rotations on both the input and output enables it to perform universal quantum computation, as it encompasses a universal set of gates. Specifically, it has been demonstrated that the QP includes the Clifford+T universal gate set~\cite{AraizaBravo2023}. Consequently, the computational complexity of a QP surpasses that of its classical counterpart, albeit with the added requirement of single-qubit rotations.

By recasting the QP into a variational circuit, we can harness its power for learning tasks, transitioning from merely using QPs as quantum circuit components to establishing them as the foundation of a novel quantum machine-learning architecture. In this approach, the rotations are treated as single-qubit unitaries, while the evolution of the Hamiltonian functions are used as entangling unitaries. The effective unitary evolution is given by
\begin{equation}
    U(\theta) = \prod_{l=1}^{L} U_{2}^{l} e^{-i \tau H_{P}^{l}} U_{1}^{l},
\end{equation}
where \( L \) signifies half the number of rotations (akin to circuit depth), $\theta$ is the set of all variational parameters, and \( \tau \) is a fixed time interval. The Hamiltonian \( H_{P}^{l} \) between rotations is given by
\begin{equation}\label{learn_perc_eq}
    H_{P}^{l} = \Omega_{o}^l \sigma_{x}^o + (-\Delta_{o}^l + \sum_{i=1}^{N} J_{i}^{l} \sigma_{z}^{i})\sigma_z^o
\end{equation}
The single-qubit rotations \( U_{1,2}^{l} \) follow an Euler representation
\begin{equation}
    U_{k}^{l} = \prod_{i=0}^{N} e^{-i \gamma_{l}^{k,i} \sigma_{z}^{i}} e^{-i \beta_{l}^{k,i} \sigma_{x}^{i}} e^{-i \alpha_{l}^{k,i} \sigma_{z}^{i}},
\end{equation}
with variational parameters \( \{ \gamma_{l}^{k,i}, \beta_{l}^{k,i}, \alpha_{l}^{k,i} \} \) for \( k = 1, 2 \).
We initialize the qubits in the state \( |0\rangle^{\otimes (N+1)} \) and measure an observable \( \hat{O} \) to produce a network output. The loss function
\begin{equation}
    L(\theta) = \frac{1}{2} \left\| \langle 0 | U^{\dagger}(\theta) \hat{O} U(\theta) | 0 \rangle - \hat{O}_{0} \right\|^2,
\end{equation}
is then minimized using gradient descent, where the gradient is given by
\begin{equation}
    \theta(t + 1) = \theta(t) - \eta \frac{\partial L}{\partial \theta(t)}.
\end{equation}
For a given dataset \( \mathcal{D} \) with data points \( x \) and labels \( \tilde{y}(x) \), the data can be loaded into the input qubits via a feature map \( x \rightarrow | \phi(x) \rangle | 0 \rangle_0 \), where $|0\rangle_0$ represents the output qubit. By measuring the observable $\hat{O}$ on the QP, we obtain
\begin{equation}
    y_{\theta}(x) = \langle 0 | \langle \phi(x) | U^{\dagger}(\theta) \hat{O} U(\theta) | \phi(x) \rangle | 0 \rangle,
\end{equation} 
rendering the loss function as
\begin{equation}
    L(\theta) = \frac{1}{2} \sum_{x \in \mathcal{D}} \left( \tilde{y}(x) - y_{\theta}(x) \right)^2.
\end{equation}

.
\vspace{-10 pt}
\subsection{Rydberg Arrays}\label{sec2c}
Rydberg atoms are typically modeled as having two states: the electronic ground state \( |g\rangle \) and the Rydberg state \( |r\rangle \), to which $|g\rangle$ can transfer through a two-photon optical transition. The atoms are positioned in a predetermined spatial arrangement, starting in a well-defined ground state. A laser light induces time evolution by coupling the atoms to their Rydberg states along the axis of the array. The final states are read out by reactivating the traps and imaging the ground-state atoms via atomic fluorescence, while Rydberg atoms are expelled~\cite{Rydberg2023}. 
Upon reaching the Rydberg state~\cite{Jaksch2000, Weimer2010}, atoms engage in strong, repulsive van der Waals interactions of frequency \(V/d^6 \), where \( d \) represents the interatomic distance. These dynamics can be described mathematically by the many-body Hamiltonian
\begin{equation}\label{rydberg_eq}
    \frac{H}{\hbar} = \sum_i \frac{\Omega_i}{2} \sigma^i_{x} - \sum_i \Delta_i n_i + \sum_{i < j} V_{ij} n_i n_j.
\end{equation}

\noindent In this equation, $n_i = |r_i\rangle \langle r_i|$, \( \Delta_i \) signifies the detuning from the Rydberg state caused by the driving laser, while \( \sigma_{x}^i \) accounts for the coupling between the ground state and Rydberg state at Rabi frequency \( \Omega_i \) (FIG.~\ref{fig:wide_figure}(a)). 

One of the key phenomena in this setup is the 'Rydberg blockade,' which prevents the simultaneous Rydberg excitation of neighboring atoms when the interatomic interactions \( V_{ij} \) surpass the effective Rabi frequency \( \Omega \). In closely-packed arrays, Rabi oscillations between the ground state and a single collective excited state occur with frequency \(\propto  \sqrt{N} \)~\cite{Dudin2012, Labuhn2016, Zeiher}. 
\vspace{-10 pt}
\section{Mapping Rydberg Hamiltonians to Perceptron Dynamics}\label{sec:mapping}

 To establish the feasibility of implementing a QP on a Rydberg array, we proceed by demonstrating how the perceptron Hamiltonian can be mapped onto that of Rydberg systems. The following derivation illustrates this mapping, providing a clear path to realizing the QP using Rydberg atoms. 

We start by defining the Rydberg state operator \( n_i \) as 
\begin{equation}
   n_i = |r_i\rangle \langle r_i| = \frac{\mathbb{I} - \sigma_z^i}{2}
\end{equation}
In the case of the QP, we restrict ourselves to configurations where such interactions only occur between the output qubit with the input qubits (for a detailed discussion on the experimental protocol and practical implementation, refer to Section~\ref{exp}), such that the index $i$ runs from $1$ to $N$ and the index $j = N + 1$, where we denote $N+1$ as $o$ for the output qubit. Likewise, as we restrict ourselves to interactions between the input qubits and the output qubit, we can simplify the notation $V_{ij}$ as $V_j$ without loss of generality. Substituting these definitions into Eq. (\ref{rydberg_eq}) we obtain:
\begin{align}\label{eq:map}
&= \sum_{i=1}^{N} \frac{\Omega_i}{2} \sigma_x^i + \sum_{i=1}^{N} \Delta_i \left( \frac{\sigma_z^i}{2} \right) \nonumber + \frac{\Omega_o}{2} \sigma_x^o + \frac{\Delta_o}{2} \sigma_z^o\\
&\quad + \sum_{i=1}^{N} \frac{V_i}{4} \left( -\sigma_z^i - \sigma_z^o + \sigma_z^i \sigma_z^o \right) \nonumber \\
&\quad + \left[ -\left( \sum_{i=1}^{N} \frac{\Delta_i}{2} \right) - \frac{\Delta_o}{2} + \left( \sum_{i=1}^{N} \frac{V_i}{4} \right) \right] \nonumber \\
&= \sum_{i=1}^{N} \frac{\Omega_i}{2} \sigma_x^i + \sum_{i=1}^{N} \left(\frac{\Delta_i}{2}- \frac{V_i}{4}\right)\sigma_z^i \nonumber \\
&\quad + \sum_{i=1}^{N} \frac{V_i}{4} \sigma_z^i \sigma_z^o + \frac{\Omega_o}{2} \sigma_x^o + \sigma_z^o \left( \frac{\Delta_o}{2} - \sum_{i=1}^{N} \frac{V_i}{4} \right) \nonumber \\
&\quad + \left[ -\left( \sum_{i=1}^{N} \frac{\Delta_i}{2} \right) - \frac{\Delta_o}{2} + \left( \sum_{i=1}^{N} \frac{V_i}{4} \right) \right].
\end{align}
To eliminate the \(\sigma_z^i\) terms, the condition 
\[
\sum_{i=1}^{N} \left(\frac{\Delta_i}{2} - \frac{V_i}{4}\right) \sigma_z^i = 0
\]
must hold, implying \(V_i = 2\Delta_i\). By comparing with Eq. (\ref{perc_eq}), we identify the following relationships:
\[
\frac{V_i}{4} = J_i, \quad \left( \frac{\Delta_o}{2} - \sum_{i=1}^{N} \frac{V_i}{4} \right) = -\Delta_o', \quad \frac{\Omega_o}{2} = \Omega_o'.
\]
The constant term \(-\Delta_o / 2\) can be omitted from the Hamiltonian as it does not influence observables. Under the condition \(V_i = 2\Delta_i\), the Hamiltonian of the Rydberg atom array can be mapped onto that of the perceptron. Given that \(V_i\) is typically around 10 MHz and the detuning \(\Delta_i\) can be adjusted to values within the same range, this mapping is practical. However, Eq. (\ref{eq:map}) includes additional terms \(\sum_{i=1}^{N} \frac{\Omega_i}{2} \sigma_x^i\) not present in Eq. (\ref{perc_eq}). These terms represent the Rabi frequencies that drive transitions between states \(|e\rangle\) and \(|g\rangle\) for the input qubits. Setting \(\Omega_i = 0\) to remove these terms would eliminate the coupling between input qubits and the driving field, effectively nullifying the laser light and rendering detuning terms meaningless. Without \(\Omega_i\), the input states would not be driven, and \(\Delta_i\) would have no functional significance. 

To preserve the interaction and maintain meaningful dynamics, we can introduce corresponding \(\frac{\Omega_i}{2} \sigma_x^i\) terms into the perceptron Hamiltonian:
\begin{equation}\label{perc_eq_mod}
   H_P = -\Delta_o' \sigma^o_z + \Omega_o' \sigma^o_x + \sum_{i=1}^{N} J_i \sigma^i_z \sigma^o_z + \sum_{i=1}^{N} \frac{\Omega_i}{2} \sigma_x^i.
\end{equation}
Including these \(\Omega_i\) terms allows for non-zero Rabi frequencies, ensuring active driving of the input qubits and preserving the relevance of the detuning terms. The perceptron model's capability to handle single-qubit rotations makes this adjustment a natural extension of the Hamiltonian. Thus, we have effectively mapped the Hamiltonian to a modified version of the perceptron model, enhancing its physical relevance and interpretability. 
\vspace{-10 pt}
\section{Experimental Realization}\label{exp}
Within arrays of Rydberg atoms, there are exciting and promising ways to experimentally encode QPs. For example, recent experiments led by Browaeys~\cite{Browaeys1, Browaeys2}, have demonstrated the controllability over ensambles of Rubidium-87 atoms which are excited to two Rydberg states (i.e., to $|s\rangle = |60S_{1/2}, m_J=1/2\rangle$ using a two-photon process, and then coherently connecting this state to $|p\rangle = |60P_{1/2}, m_J=-1/2\rangle$ via a microwave field). In this case, two atoms in Rydberg states interact via dipole-dipole or quadruple interactions. For example, when two atoms, labeled $1$ and $2$, are in the same state, their interactions follows a Van der Waals potential $H_{int}^6=C_6/R_{1,2}^6$ where the value of $C_6$ depends on the state of the atoms. Meanwhile, atoms in opposite Rydberg states interact via a flip-flop Hamiltonian
\begin{equation*}
    H^3_{int} = \frac{d^2(3\cos^2\theta_{12}-1)}{R_{12}^3}\left(s_1^\dagger p_2 + p_1^\dagger s_2\right)
\end{equation*}
where $d$ is the transition dipole moment between the two Rydberg levels, $\theta_{ij}$ is the angle of $R_{ij}$ with respect to the quantization axis defined by an external magnetic field, and $s_i^\dagger, p_i^\dagger$ are the creation operators of each Rydberg state in atom $i$. By defining the Pauli matrices $\sigma_{x,y}^i$ in the $\{|s\rangle, |p\rangle\}$ subspace, it can be shown that $(s_1^\dagger p_2 + \text{h.c.}) \propto \sigma^1_x\sigma^2_x+\sigma^1_y\sigma^2_y$. Moreover, by arranging atoms $i_1, i_2$ with $\theta_{i_1i_2}=\arccos\left(1/\sqrt{3}\right)$, these atoms do not interact via $H_{int}^3$. Thus, atoms arranged in a line for a 2D setup (or on a plane for a 3D-setup) $\arccos\left(1/\sqrt{3}\right)$ away from the quantization axis only interact via $H^{6}_{int}$ while they interact via $H^{3}_{int}$ with an atom outside this line (or plane). In the case that the flip-flop terms are much larger than the Van der Waals interactions as is the case for these specific states, we obtain the modified perceptron interaction 
\begin{equation}
    H_{P} = \sum_{i}J_{i}\left(\sigma^i_x\sigma_x^o+\sigma^i_y\sigma_y^o\right)
\end{equation}
which is a generalization of the original QP proposal.\par 
Another exciting experimental platform to realize QPs with Rydberg atoms is that of dual-species arrays. For example, a team led by Bernien has recently demonstrated configurable arrays of Rubidium-87 and Cesium-133 atoms~\cite{Bernien1}. While current experiments operate in the regime where Cs-Rb cross-talk is negligible, theoretical calculations~\cite{Beterov2015} show that in some cases the inter-species interactions dominate intra-species interactions. Take, for example, the case of $|r\rangle = |81S_{1/2}\rangle$ reported in Ref.~\cite{Beterov2015}. In this case, the inter-species interaction $V_{\text{Cs,Rb}}$ and the intra-species one $V_{\text{Cs}}, V_{\text{Rb}}$ are related via the ratios $V_{\text{Cs}}/V_{\text{Rb}}\approx 0.77$ and $V_{\text{Cs,Rb}}/V_{\text{Rb}}=21.9$ which means that the inter-species interaction is about two orders of magnitude greater than the intra-species one. As a result, for times such that $V_{\text{Cs,Rb}}\tau\approx 1$, the dual-species arrays can be regarded as encoding the input qubits in one species and the output qubits in another. Since each species can be read-out seperately, it also means that measurement of the output qubits are a non-demolition measurement of the input qubits. \par 
\begin{figure}[ht]
    \centering
    \includegraphics[scale=0.55]{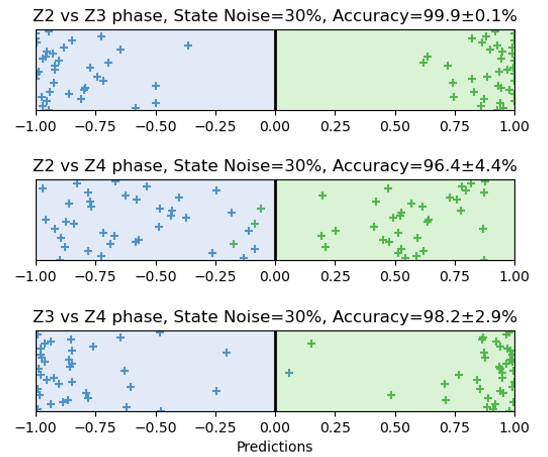}
    \caption{Classification of Z2, Z3 and Z4 phases with 30\% probability amplitude error. Each plot shows the clear separation between the respective phases}
    \label{phase}
\end{figure}

\section{Phase Classification}\label{sec_phase}

We explore the use of QPs in QML tasks, specifically targeting a germane task for Rydberg systems: phase classification. Rydberg atoms, due to the Rydberg blockade effect can lead to different quantum phases~\cite{Pohl2010} depending on the lattice configurations and interaction range. These phases arise from the interplay of coherent coupling and interatomic interactions, leading to distinct spatial ordering and symmetry breaking. 

Our evaluation consisted of three primary tasks: (1) assessing the QP's accuracy in classifying states within the Z2, Z3 and Z4 phases, defined as
\begin{center}
Z2 : \(|10101010\rangle, |01010101\rangle\)\\
Z3 : \(|10010010\rangle, |01001001\rangle, |00100100\rangle\)\\
Z4 : \(|10001000\rangle, |01000100\rangle, |00100010\rangle, |00010001\rangle\)
\end{center} (2) examining its performance in classifying states subjected to varying noise levels, and (3) analyzing QP's ability to classify states in the disordered and Z2 phases under different noise levels. FIG.~\ref{fig:wide_figure}(b) illustrates the Z2 and Z3 phases with different broken symmetries.

To evaluate the robustness of our approach, we introduced a 30\% probability amplitude error margin in qubit state determination. Specifically, we sampled a random number $r$ from a uniform distribution over the interval [0.7,1), and utilized its square root to determine the coefficients for the quantum states. The state \( |\psi\rangle \) can thus be expressed as \( |\psi\rangle = \sqrt{r}|0\rangle + \sqrt{1-r}|1\rangle \). This method introduced a controlled level of noise into the system, replacing the ideal \( |0\rangle \) and \( |1\rangle \) states with these noisy variants. We generated a chain of 8 input qubits in either the Z2, Z3 or Z4 phase, with each qubit subject to the previously described noise. We then classified these inputs using a QP to distinguish between Z2/Z3, Z2/Z4, and Z3/Z4.
For brevity and without loss of generality, we here discuss the classification of Z2 and Z3 states. The classification accuracies for all three phase pairings are shown in FIG.~\ref{phase}.

Our training and test datasets each comprise 36 states of the Z2 type, labeled as -1, and 36 states of the Z3 type, labeled as +1. These input states are processed through a series of QPs with \(\tau\) values ranging from 0.1 to 1.0 in increments of 0.1 and $L=2$. For each QP, we measure the expectation value of the output qubit and store it in a vector \( r(\psi) \). In this evolution, we replace Eq. (\ref{learn_perc_eq}) by the Rydberg Hamiltonian (Eq. (\ref{rydberg_eq})) subjected to the conditions detailed in Section \ref{sec:mapping}, to ensure that it exhibits perceptron dynamics. 
During training, the output \( y(\psi) \) is computed using the function \( \tanh(w \cdot r(\psi)) \), where \( w \) is a vector of optimization parameters. The loss function minimizes over both the QP parameters (\(\alpha, \beta, \gamma\)) and the vector \( w \). We employ the mean squared error (MSE) loss to compare \( y(\psi) \) with the true labels across the training dataset. The Adam optimizer is utilized in our simulations to update the parameters efficiently. After training, the test dataset is used to evaluate the classification accuracy. Despite the stochasticity, the QP was able to classify between these phases with high accuracies as shown in FIG.~\ref{phase}. The high classification accuracy, even in the presence of considerable noise, suggests that QPs can be reliably used in environments where quantum states are prone to errors.
\begin{figure}
    \centering
    \includegraphics[scale=0.55]{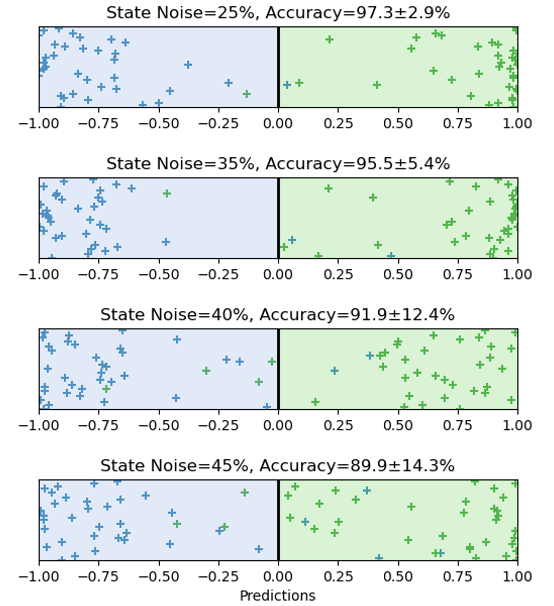}
    \caption{Classification accuracy of the QP for Z2 and Z4 phases under varying probability amplitude error levels. As noise increases from 25\% to 45\%, accuracy decreases, highlighting the impact of noise on classification performance.}
    \label{fig:phase_noise}
\end{figure}
We further investigated the classification accuracy of the QP under varying levels of state noise. Specifically, we examined its performance in distinguishing between the Z2 and Z4 phases using the procedure described earlier. To comprehensively evaluate the perceptron's robustness, we tested state error rates of 25\%, 35\%, 40\%, and 45\%.

As anticipated, our findings revealed a clear trend: as the level of state noise increased, the classification accuracy correspondingly decreased. This inverse relationship highlights the sensitivity of the QP's performance to the quality of the input quantum states. The detailed results of this analysis are presented in FIG.~\ref{fig:phase_noise}, which illustrates how the perceptron’s accuracy degrades with increasing noise levels. Despite this decline, it is worth noting that the QP remains remarkably robust overall, maintaining high accuracy even in the presence of substantial noise.

\begin{figure}
    \centering
  \includegraphics[scale=0.55]{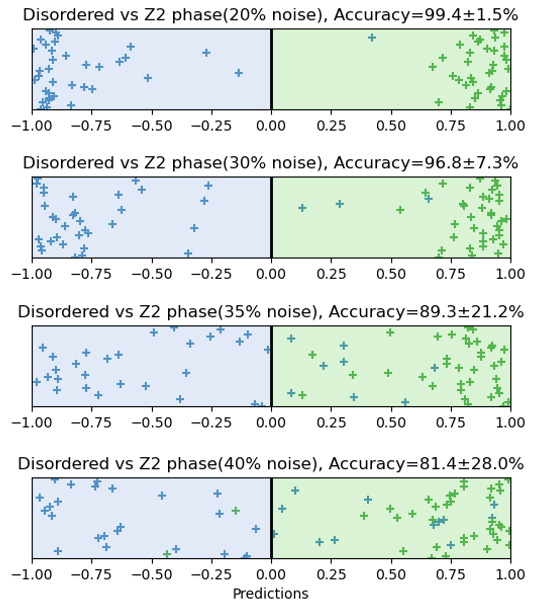}
    \caption{Classification accuracy of the QP for disordered and Z2 phase under varying levels of probability amplitude error.}
    \label{fig:phase_disorder}
\end{figure}

For our final test, we generated arbitrary states for the disordered phase and compared them to Z2 phase states under varying levels of probability amplitude error, as summarized in FIG. \ref{fig:phase_disorder}. To construct these disordered phase states, we sampled a random number \( r \) from a uniform distribution over the interval [0, 1), and used its square root to determine the coefficients for the quantum states, resulting in the form $|\psi\rangle = \sqrt{r}|0\rangle + \sqrt{1-r}|1\rangle$. As in the previous tests, we observed that higher noise levels led to a reduction in classification accuracy. Despite the randomness of the disordered states, the QP maintained robust performance, underscoring its sensitivity to noise while still accurately distinguishing between states.

\vspace{-15pt}
\section{Expanding to a Two-Output Qubit Architecture}\label{sec2oq}

A multi-layer perceptron (MLP) is an artificial neural network made up of multiple layers of neurons~\cite{bishop1995neural, hornik1991approximation}. The first step in constructing such a network by stacking perceptrons is to incorporate multiple output qubits. This enhancement allows for evaluating the model's performance in multi-class classification, where each output qubit independently interacts with all input qubits. In our approach, we extend the current perceptron model to include two output qubits (see Section~\ref{exp}). FIG.~\ref{fig:2oq_circuit} shows a schematic of a QP with two output qubits as well as its circuit representation.
The implementation of this architecture using Rydberg atoms is achievable and can be demonstrated through similar approaches as summarized below.
Eq. (\ref{perc_eq}) can be modified to include terms for both the output qubits and now evolves under the Hamiltonian:
\begin{equation}
\label{2oq_perc}
\begin{aligned}
H_P = & -\Delta_{o1}' \sigma_z^{o1} + \Omega_{o1}' \sigma_x^{o1} + \sum_{i=1}^{N} J^1_i \sigma_z^i \sigma_z^{o1} \\
& -\Delta_{o2}' \sigma_z^{o2} + \Omega_{o2}' \sigma_x^{o2} + \sum_{i=1}^{N} J^2_i \sigma_z^i \sigma_z^{o2}
\end{aligned}
\end{equation}
where $o_1$ and $o_2$ denote output qubits 1 and 2, while $J^1_i$ and $J^2_i$ denote the interaction potentials of the input qubits with $o_1$ and $o_2$, respectively.
To implement the two-output qubit perceptron using arrays of Rydberg atoms, we need to demonstrate the mapping between Eq. (\ref{rydberg_eq}) and Eq. (\ref{2oq_perc}). We apply the same methodology as described in Section~\ref{sec:mapping}.

The Rydberg Hamiltonian can be written as:
\begin{equation}\label{2oq_rydberg}
\begin{aligned}
\frac{H}{\hbar}=&\sum_i \frac{\Omega_i}{2} \sigma_x^i - \sum_i \Delta_i n_i + \sum_{i<j} V_{ij} n_i n_j \\
&=\frac{\Omega_{o1}}{2} \sigma_x^{o1} + \frac{\Omega_{o2}}{2} \sigma_x^{o2} + \sum_{i=1}^{N} \frac{\Omega_i}{2} \sigma_x^i - \Delta_{o1} n_{o1}\\
& - \Delta_{o2} n_{o2}-\sum_{i=1}^{N} \Delta_i n_i + \sum_{i=1}^{N} V_{i} n_i n_{o1} + \sum_{i=1}^{N} V'_{i} n_i n_{o2} \\
\end{aligned}
\end{equation}
where \( i \in \{1, \ldots, N\} \) represents the input qubits, with \( i = N+1 \) corresponding to $o_1$ and \( i = N+2 \) corresponding to $o_2$. All interactions between input qubits, as well as between the output qubits, have been removed to align with the perceptron architecture. Additionally, all $\Omega_i$ values can be dealt with the same way as done in Section \ref{sec:mapping} by introducing corresponding $\frac{\Omega_i}{2} \sigma_x^i$ terms in Eq. \ref{2oq_perc}. Eq. (\ref{2oq_rydberg}) can be manipulated and expressed as the following where \( n_i = \frac{\mathbb{I} - \sigma_z}{2} \):
\begin{equation}
    \begin{aligned}
        H/\hbar = &\sigma_z^i \sum_{i=1}^{N}(\frac{\Delta_i}{2} - \frac{V_{i}}{4} - \frac{V'_{i}}{4}) + \sigma_z^{o1} (\frac{\Delta_{o1}}{2} - \sum_{i=1}^{N} \frac{V_i}{4}) \\
        & + \sigma_z^{o2} (\frac{\Delta_{o2}}{2} - \sum_{i=1}^{N} \frac{V'_i}{4}) + \sum_{i=1}^{N} \frac{V_i \sigma_z^i \sigma_z^{o1}}{4} \\
        & + \sum_{i=1}^{N} \frac{V'_i \sigma_z^i \sigma_z^{o2}}{4} + \frac{\Omega_{o1}}{2} \sigma_x^{o1} + \frac{\Omega_{o2}}{2} \sigma_x^{o2} + \sum_{i=1}^{N} \frac{\Omega_i}{2} \sigma_x^i
    \end{aligned}
\end{equation}

On comparing terms with the modified perceptron Hamiltonian, we obtain the following conditions:
\begin{equation}
\frac{\Delta_{o1}}{2} - \sum_{i=1}^{N} \frac{V_i}{4} = \Delta'_{o1};   \frac{\Delta_{o2}}{2} - \sum_{i=1}^{N} \frac{V'_i}{4} = \Delta'_{o2}
\end{equation}
\begin{equation}
\frac{V_i}{4} = J_i^1, \frac{V'_i}{4} = J_i^2, \frac{\Omega_{o1}}{2} = \Omega'_{o1}, \frac{\Omega_{o2}}{2} = \Omega'_{o2}
\end{equation}
\begin{equation}
    2\Delta_i = V_i + V'_i
\end{equation}

To evaluate the performance of this perceptron with two output qubits, we revisit entanglement classification. The most well-known types of entanglement classification are local unitary (LU), local operations and classical communication (LOCC), and stochastic local operations and classical communication (SLOCC)~\cite{Ghahi2016}. In a two-qubit system, entanglement is classified as either separable or entangled. However, as the number of qubits increases, the classification becomes more intricate. For instance, in a three-qubit system, SLOCC requirements result in a classification that includes one class of separable state, three classes of biseparable states, and two classes of genuinely entangled states (GHZ and W). For systems with \( n \)-qubits where \( n \geq 4 \), SLOCC classification involves an infinite number of classes~\cite{Assadi2016}. For simplicity, we limit the number of states and classes considered for input qubits when \(n=10\).

\begin{table}
\centering
\begin{tabular}{|c|>{\centering\arraybackslash}p{6cm}|}
\hline
Class & State \\
\hline
Separable & \( \left|00000000\right\rangle \) \\
\hline
\multirow{7}{*}{Tri-Separable} & \( \left|00000000\right\rangle + \left|00111111\right\rangle \) \\
 & \( \left|00000000\right\rangle + \left|10011111\right\rangle \) \\
 & \( \left|00000000\right\rangle + \left|11001111\right\rangle \) \\
 & \( \left|00000000\right\rangle + \left|11100111\right\rangle \) \\
 & \( \left|00000000\right\rangle + \left|11110011\right\rangle \) \\
 & \( \left|00000000\right\rangle + \left|11111001\right\rangle \) \\
 & \( \left|00000000\right\rangle + \left|11111100\right\rangle \) \\
\hline
\multirow{7}{*}{Bi-Separable} & \( \left|10000000\right\rangle + \left|11111111\right\rangle \) \\
 & \( \left|01000000\right\rangle + \left|11111111\right\rangle \) \\
 & \( \left|00100000\right\rangle + \left|11111111\right\rangle \) \\
 & \( \left|00010000\right\rangle + \left|11111111\right\rangle \) \\
 & \( \left|00001000\right\rangle + \left|11111111\right\rangle \) \\
 & \( \left|00000100\right\rangle + \left|11111111\right\rangle \) \\
 & \( \left|00000010\right\rangle + \left|11111111\right\rangle \) \\
\hline
\multirow{2}{*}{Inseparable} & \(\text{GHZ: } \left|00000000\right\rangle + \left|11111111\right\rangle \) \\
 & \(\text{W: } \left|10000000\right\rangle + \left|01000000\right\rangle + \left|00100000\right\rangle + \left|00010000\right\rangle +\) \\
 & \(\left|00001000\right\rangle + \left|00000100\right\rangle + \left|00000010\right\rangle + \left|00000001\right\rangle\) \\
\hline
\end{tabular}
\caption{Summary of states and classes considered for the 8-input qubit system}
\label{table2}
\end{table}
We generate a dataset \( D \) comprising four types of quantum states based on their separability and evaluate whether our perceptron can accurately classify them. The dataset includes states mixed in equal proportions from each class, labeled as \([-1, -1]\), \([-1, 1]\), \([1, -1]\), and \([1, 1]\). These states are subjected to probability amplitude error, the amount of which can be quantified using fidelity. Quantum fidelity is a widely recognized metric for assessing the similarity between quantum states. For our purposes, we maintain an average fidelity of 0.95 (ranging from 0.93 to 0.97) between the noisy states and their corresponding pure states. To generate the quantum states in our dataset \( D \), we began by generating a random number \( r \) between 0.99 and 1 using a uniform distribution. The square root of \( r \) was then used to determine the coefficients for the quantum state \( |\psi\rangle \), which was expressed as \( |\psi\rangle = \sqrt{r}|0\rangle + \sqrt{1 - r}|1\rangle \), following the same approach as in the phase classification (Section~\ref{sec_phase}). The fidelity of the state was calculated by taking the squared inner product between the pure state and the state subjected to probability amplitude error. If the fidelity fell within the specified range, the state was added to the dataset; otherwise, a new state was generated. These states are then processed through a series of QPs with a depth of $L=4$. The outputs were determined by post-processing the final measurements, analogous to the procedure used in the single output qubit case. We compared these outputs to the true labels to calculate the loss function and optimized the QP parameters and weights using the Adagrad optimizer.

Our objective was to classify quantum states into four categories: separable, inseparable, bi-separable, and tri-separable, as summarized in TABLE~\ref{table2}. 
\begin{figure}[h]
    \centering
    \includegraphics[scale = 0.5]{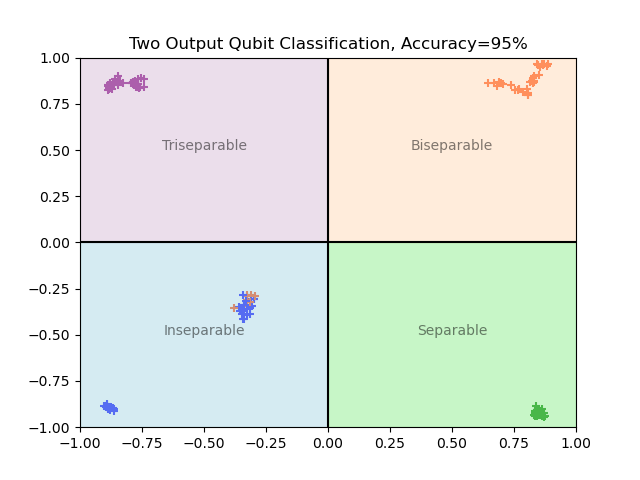}
    \caption{Classification results for a 10-qubit system into four categories: separable, inseparable, bi-separable, and tri-separable states. The QP achieved 95\% accuracy on the test dataset, with each quadrant representing a distinct class of quantum states.}
    \label{fig:2oq}
\end{figure}

An average fidelity of 0.95 (ranging from 0.93 to 0.97) was maintained between the noisy states and their corresponding pure statese The QP demonstrated remarkable performance, achieving 95\% accuracy on the test dataset which is illustrated in FIG.~\ref{fig:2oq}. Each quadrant corresponds to a different category of quantum state, and the distinct clustering of data points within each quadrant indicates the successful classification by the perceptron. Similar tests for 5- and 6- qubit systems yielded consistent results, as detailed in Appendix~\ref{2oq}.

\section{UNIVERSAL APPROXIMATION THEOREM AND ERROR BOUNDS}\label{errorbound}
In recent years, the theoretical underpinnings of quantum neural networks have been bolstered by developments in universal approximation theorems~\cite{Perez-Salinas2020, Perez-Salinas2021, Schuld2021}, which establish that quantum circuits can arbitrarily approximate complex-valued continuous functions. While these theorems form the bedrock of our understanding, they often lack precise quantification of approximation errors—a gap that is crucial for both theoretical completeness and practical implementations. Recent advancements have addressed this shortcoming by formulating exact error bounds for quantum neural network approximations. Specifically, it has been demonstrated by Gonon and Jacquier~\cite{error} that a parameterized quantum circuit can approximate continuous functions bounded in \( L_1 \) norm up to an error of order $n^{-\frac{1}{2}}$, where the number of qubits scales logarithmically with $n$. More precisely, it was shown that a quantum neural network with \( O(\epsilon^{-2}) \) weights and \( O(\lceil \log_2 (\epsilon^{-1}) \rceil) \) qubits suffices to achieve accuracy \( \epsilon > 0 \) when approximating functions with integrable Fourier transform.

According to Theorem 2.4 in Ref.~\cite{error}, for any constant \( R > 0 \) and \( n \in \mathbb{N} \), there exists \( \theta \in \Theta \) such that
\[
\left( \int_{\mathbb{R}^d} \left| f(x) - f_{R,n,\theta}(x) \right|^2 \mu(dx) \right)^{1/2} \leq \frac{L_1[\hat{f}]}{\sqrt{n}}.
\]
In this equation, \( \theta \) represents the set of parameters that define the rotation angles in the quantum circuit, \( f(x) \) is the target function to be approximated, and \( f_{R,n,\theta}(x) \) is the function that the quantum circuit approximates. The term \( L_1[\hat{f}] \) denotes the \( L_1 \)-norm of the Fourier transform of the function \( f \). The input \( x \) lies in the space \( \mathbb{R}^d \), and \( n = 2^{\mathbf{n}}/4 \), where \( \mathbf{n} \) is the number of qubits used in the quantum circuit.

It was also demonstrated that a similar result holds when the original parameterized quantum circuit is replaced by a reservoir quantum circuit, where all but the final unitary operators are randomized and fixed, drawing parallels to classical random feature networks and reservoir computing. This confluence of error bounds in classical and quantum settings not only reinforces the computational universality of quantum neural networks but also provides a roadmap for their efficient implementation.
We demonstrate that the specified parameterized quantum circuit (Appendix~\ref{qc_arch_error}) is implementable using a QP (Appendix~\ref{appB}), and thus, the established error bounds (Appendix~\ref{sec:approx_error}) naturally extend to the perceptron model as well. 
This equivalence not only validates the QP's capability for approximating classical functions, but also provide precise error margins for such approximations, reinforcing their role as reliable building blocks for scalable quantum neural networks. 

\section{Conclusion}\label{conclusion}

In this manuscript, we present the implementation of QPs using Rydberg atom arrays by mapping Rydberg Hamiltonians onto perceptron dynamics. QP dynamics on Rydberg arrays are demonstrated to be effective in classifying phases with different broken symmetries, as well as in entanglement classification in the presence of varying levels of probability amplitude error. Experimental strategies for encoding QPs on arrays of Rydberg atoms are examined, focusing on the necessary interactions between input qubits and the output qubit while preventing interactions among input qubits. Two approaches are considered: single-species arrays, utilizing Van der Waals and flip-flop Hamiltonians, and dual-species arrays, where different atomic species are used to encode input and output qubits, enabling strong inter-species interactions and non-demolition measurements.  Additionally, the approximation capabilities of variational quantum circuits are extended to QPs, showing that they can approximate continuous functions with an error scaling as $n^{-\frac{1}{2}}$, where the number of qubits required increases logarithmically with $n$. This ensures that no curse of dimensionality occurs.

The QP architecture, with its simple and modular design, represents a crucial advancement toward scalable QML models. Realizing QPs on quantum hardware (Section~\ref{sec:mapping}) is the next critical step, a challenge that we address in this manuscript. In Section~\ref{sec2oq}, the perceptron is further explored by expanding the QP model to include two output qubits, positioning its realization within deep neural network structures and enhancing its potential for multi-class classification tasks. Thus, this work lays the foundation for future experimental efforts in realizing this model for practical QML applications.

Further research in this area could include experimental validation of our findings, which would substantiate the practical applicability of the QP model. Additionally, the model could be enhanced by incorporating multiple output qubits or stacking QPs, potentially increasing its computational power and versatility. Moreover, integrating quantum reservoir computing with QPs could significantly refine the gradient-based learning process, reducing model training to a simple linear regression problem that uses post-processed QP measurements and is carried out on classical hardware~\cite{martinez2020information, bravo2022quantum, mujal2021opportunities, vanderSande2017advances}. This technique is particularly useful for noisy quantum hardware~\cite{ghosh2021,suzuki2022}.

\section{Acknowledgements}
We acknowledge the support from various institutions and programs. IA is grateful for the Caltech Summer Undergraduate Fellowship. RAB acknowledges support from the National Science Foundation (NSF) Graduate Research Fellowship under Grant No. DGE1745303. SFY thanks the NSF through a Cornell HDR (OAC-2118310) and the CUA PFC (PHY-2317134) grant. A. Anandkumar’s work is supported in part by the Bren endowed chair, the ONR (MURI grant N00014-18-12624), and the AI2050 Senior Fellow Program at Schmidt Sciences.

\bibliography{lib}
\onecolumngrid
\appendix
\section{Multiclass Classification}\label{2oq}

\begin{table*}[htbp]
\centering
\caption{Summary of classes and states for different number of input qubits}
\begin{tabularx}{\textwidth}{|>{\centering\arraybackslash}m{4cm}|>{\centering\arraybackslash}m{5cm}|>{\centering\arraybackslash}X|}
\hline
\textbf{No. of Qubits} & \textbf{Class} & \textbf{State} \\
\hline
\multirow{6}{*}{3} 
 & GHZ & \( \left|000\right\rangle + \left|111\right\rangle \) \\
 \cline{2-3}
 & W & \( \left|001\right\rangle + \left|010\right\rangle + \left|100\right\rangle \) \\
 \cline{2-3}
 & Separable & \( \left|000\right\rangle \) \\
 \cline{2-3}
 & \multirow{3}{*}{Bi-Separable} & \( \left|001\right\rangle + \left|111\right\rangle \) \\
 & & \( \left|010\right\rangle + \left|111\right\rangle \) \\
 & & \( \left|100\right\rangle + \left|111\right\rangle \) \\
\hline
\multirow{10}{*}{4} 
 & \multirow{2}{*}{Inseparable} & \(\text{GHZ: } \left|0000\right\rangle + \left|1111\right\rangle \) \\
 & & \(\text{W: } \left|0001\right\rangle + \left|0010\right\rangle + \left|0100\right\rangle + \left|1000\right\rangle \) \\
 \cline{2-3}
 & Separable & \( \left|0000\right\rangle \) \\
 \cline{2-3}
 & \multirow{4}{*}{Bi-Separable} & \( \left|1000\right\rangle + \left|1111\right\rangle \) \\
 & & \( \left|0100\right\rangle + \left|1111\right\rangle \) \\
 & & \( \left|0010\right\rangle + \left|1111\right\rangle \) \\
 & & \( \left|0001\right\rangle + \left|1111\right\rangle \) \\
 \cline{2-3}
 & \multirow{3}{*}{Tri-Separable} & \( \left|0000\right\rangle + \left|1100\right\rangle \) \\
 & & \( \left|0000\right\rangle + \left|1001\right\rangle \) \\
 & & \( \left|0000\right\rangle + \left|0011\right\rangle \) \\
\hline
\end{tabularx}
\label{table:ent}
\end{table*}

 The dataset's input states considered for 5- and 6- qubit systems are summarized in TABLE~\ref{table:ent}. Remarkably, the model achieved 100\% accuracy on the test datasets for both systems by the end of training.  This high accuracy can be attributed to the relatively small system sizes compared to the model's complexity and the size of the training dataset. Smaller systems exhibit nearly deterministic behavior, making them fully learnable toy models.

\section{Quantum Circuit Architecture}\label{qc_arch_error}
 As outlined in Section 2 of Ref.~\cite{error}, we begin by introducing the parameterized quantum circuit, which is defined by a set of hyperparameters or rotation angles. For \( n \in \mathbb{N} \), let the weight vectors \( \mathbf{a} = (a_1, \ldots, a_n) \) be in \( (\mathbb{R}^d)^n \), \( \mathbf{b} = (b_1, \ldots, b_n) \) be in \( \mathbb{R}^n \), and \( \boldsymbol{\gamma} = (\gamma_1, \ldots, \gamma_n) \) lie in \( [0, 2\pi]^n \). For an input vector \( \mathbf{x} = (x_1, \ldots, x_d) \) in \( \mathbb{R}^d \), we define the following gates acting on a single qubit:
 \begin{equation}
\mathrm{U}_1^{(i)}(\mathbf{a}^i,b^i,\mathbf{x}):=\mathrm{H} \, \mathrm{R_z}(-b^i) \, \mathrm{R_z}(-a_d^i x_d) ... \mathrm{R_z}(-a_1^i x_1) \, \mathrm{H}
 \end{equation}
\begin{equation}
    \mathrm{U}_2^{(i)}(\gamma^i):=\mathrm{R_y}(\gamma^i)
\end{equation}
We then define 
\[
\theta = \left( a^{(i)}, b^{(i)}, \gamma^{(i)} \right)_{i=1,\ldots,n} \in \Theta := ( \mathbb{R}^d \times \mathbb{R} \times [0, 2\pi] )^n
\]
and the block matrix $\mathrm{U} (\mathbf{\theta},\mathbf{x})\in\mathbb{C}^{N \times N}$ constructed using the single qubit gates $\mathrm{U}_1^{(i)}$ and $\mathrm{U}_2^{(i)}$. It can be viewed as a gate operating on $\mathbf{\mathrm{n}}=\log_2(4n)$ qubits, representing our system size, with $N= 4n= 2^\mathbf{\mathrm{n}}$.
\[
\mathrm{U}(\mathbf{\theta},\mathbf{x}):=
    \begin{bmatrix}
    \mathrm{U_1^{(1)} \otimes U_2^{(1)}} & \mathbf{0_{4\times 4}} & \cdots & \mathbf{0_{4\times 4}} \\
     \mathbf{0_{4\times 4}} & \mathrm{U_1^{(2)} \otimes U_2^{(2)}}  & ... & \mathbf{0_{4\times 4}} \\
     \vdots & \ddots & \vdots & \vdots\\
      \mathbf{0_{4\times 4}} & \cdots & \mathbf{0_{4\times 4}} & \mathrm{U}_1^{(n)} \otimes \mathrm{U}_2^{(n)} \\
    \end{bmatrix}.
\]
We also define $V \in \mathbb{C}^{N \times N}
$ as any unitary matrix that maps the state $\left| 0 \right\rangle^{\otimes \mathrm{n}}$ to the state $ \left| \psi_i \right\rangle = \sqrt{\frac{1}{n}} \sum_{i=0}^{n-1} \left| 4i \right\rangle$. Here, \( |4i\rangle \) denotes the qubit state corresponding to the integer representation \( 4i \), analogous to how \( |000\rangle \) corresponds to \( |0\rangle \) and \( |101\rangle \) corresponds to \( |5\rangle \). 

For \( n \in \mathbb{N} \) and parameter \( \theta \in \Theta \), we introduce the unitary operator \( \mathrm{C_n}(\theta, \mathbf{x}):= \mathrm{U}(\mathbf{\theta}, \mathbf{x})\mathrm{V} \). This operator, which acts on \( \mathbf{\mathrm{n}} \) qubits, represents the implementation of the variational quantum circuit and the final state of the $\mathbf{\mathrm{n}}$ qubits is measured after applying the gates $\mathrm{V}$ and $\mathrm{U}(\theta,\mathbf{x})$ consecutively. The possible states that we could measure are \( 0, 1, \ldots, N - 1 \). Thus, for \( m \in \{0, 1, 2, 3\} \), we define
\[ P_n^m := P\left( \mathrm{C_\mathrm{n}}(\theta, \mathbf{x}) |0\rangle^{\otimes \mathrm{n}} \in \{m, 4 + m, \ldots, 4(n - 1) + m\} \right) \]
where \( P_n^m \) denotes the probability that the state of the \(\mathrm{n}\)-qubit system, after being processed by the quantum circuit \( \mathrm{C_n}(\theta, \mathbf{x}) \) starting from the initial state \( |0\rangle^{\otimes \mathrm{n}} \), is found in the set of states \(\{m, 4 + m, \ldots, 4(n - 1) + m\}\). This allows us to analyze the distribution of measured states and evaluate the performance of the quantum circuit under different configurations and parameter settings.
As demonstrated in Section 2.3 of Ref. \cite{error}, for any choice of weights \(\theta = (a_i, b_i, \gamma_i)_{i=1,\ldots,n}\), the function \(g_{R,n,\theta} : \mathbf{x} \mapsto \frac{1}{n} \sum_{i=1}^{n} R \cos(\gamma_i) \cos(l^i(\mathbf{x}))\), where \(l^i(\mathbf{x}) = b^i + a^i \cdot \mathbf{x}\), aligns with the expression for the map \(f_{R,n,\theta}\) given by
\begin{equation}\label{error_map}
    f_{R,n,\theta}(\cdot) := R - 2R[P_n^1(\theta, \cdot) + P_n^2(\theta, \cdot)],
\end{equation}
which can thereby be represented as
\begin{equation}
    f_{R,n,\theta}(\mathbf{x}) = \frac{1}{n}\sum_{i=1}^{n} R \cos(\gamma_i) \cos(l^i(\mathbf{x})).
\end{equation}
This expression denotes the final function that the quantum circuit yields and is particularly significant when compared to the target function that needs to be approximated (details in Appendix~\ref{sec:approx_error}) . Functions of this type have been shown to effectively approximate continuous, integrable functions $f$ with integrable Fourier transforms.

\section{Realizing the circuit using QPs}\label{appB}
In demonstrating the QP's ability to implement the prescribed quantum circuit, we leverage its inherent potential for universal quantum computation. The first step in doing so is to demonstate how we can transition from the initial state \( \left| 0 \right\rangle^{\otimes \mathrm{n}} \) to the targeted state \( \left| \psi_i \right\rangle = \sqrt{\frac{1}{n}} \sum_{i=0}^{n-1} \left| 4i \right\rangle \)~\cite{error}, which can be accomplished by applying the Hadamard gate to the first \( \mathrm{n}-2 \) qubits. As an illustrative example, consider the state \( \left| 0000 \right\rangle \). Application of the Hadamard gate to the first qubit transforms it into \( \frac{1}{\sqrt{2}} \left( \left| 1000 \right\rangle + \left| 0000 \right\rangle \right) \). Subsequent application of the Hadamard gate to the second qubit yields \( \frac{1}{2} \left( \left| 0000 \right\rangle + \left| 0100 \right\rangle + \left| 1000 \right\rangle + \left| 1100 \right\rangle \right) \), which aligns with our objective state.
For an \( \mathrm{n} \)-qubit state \( |0\rangle^{\otimes \mathrm{n}} \), applying the Hadamard gate to the first \( \mathrm{n}-2 \) qubits results in the unitary transformation represented by \( \mathrm{V}=\mathrm{H}^{\otimes (\mathrm{n}-2)} \otimes \mathrm{I} \otimes \mathrm{I} \), where \( \mathrm{I} \) denotes the identity matrix for the remaining \( 2 \) qubits. 
The QP can naturally implement elementary single-qubit rotations—namely \( R_x \), \( R_y \), and \( R_z \)—as well as foundational gates like the Hadamard and Identity gates. By utilizing the two output qubit perceptron model, we can transition from the initial state \( \left| 0 \right\rangle^{\otimes \mathrm{n}} \) to the targeted state, and in turn successfully realise the gate $\mathrm{V}$, by applying Hadamard gates only to the input qubits.

\begin{figure}[h]
    \centering
    \includegraphics[scale=0.4]{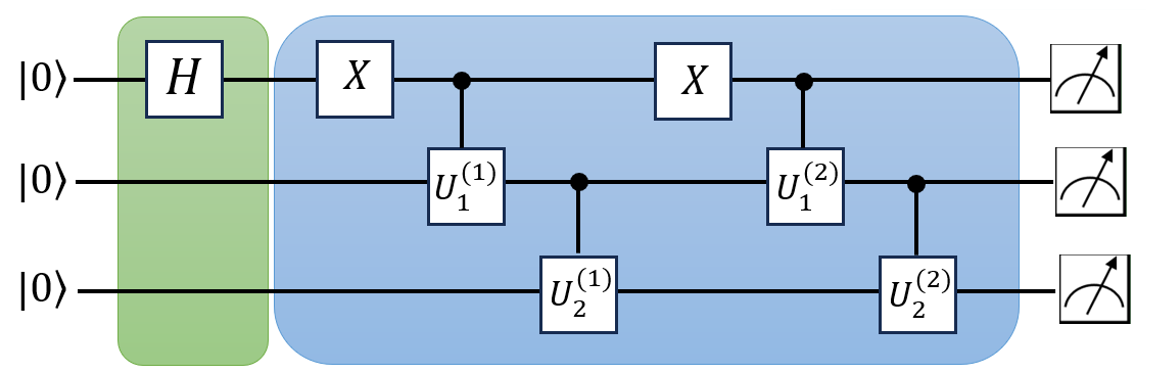}
    \caption{Quantum circuit $\mathrm{C_n}(\theta,\mathbf{x})$ implementing the \(\mathrm{V}\)(green) and \(\mathrm{U}\)(blue) gates for a 3-qubit system. The input qubit controls the application of single-qubit rotations \(\mathrm{U_1^{(i)}}\) and \(\mathrm{U_2^{(i)}}\) on the two output qubits. If the input qubit is in state zero, \(\mathrm{U_1^{(1)}}\) and \(\mathrm{U_2^{(1)}}\) are applied; if the input qubit is in state one, \(\mathrm{U_1^{(2)}}\) and \(\mathrm{U_2^{(2)}}\) are applied.
}
    \label{fig:error_circuit}
\end{figure}

To realize gate \( \mathrm{U(\theta,\mathbf{x})} \) using our QP, it's essential to understand its fundamental mechanism based on how it has been defined: it applies specific single-qubit rotations, $\mathrm{U_1^{(i)}}$ and $\mathrm{U_2^{(i)}}$, to the last two qubits, contingent on the state of the preceding $\mathbf{\mathrm{n}}-2$ qubits.  For implementing this within the QP framework, we can once again use the two output qubit model, where \(\mathrm{U_1^{(i)}}\) and \(\mathrm{U_2^{(i)}}\) specify the rotations applied to the output qubits based on the state of the input qubits. For instance, in a 3-qubit system, when the input qubit is set to the zero state, the operation $\mathrm{U_1^{(1)} \otimes U_2^{(1)}}$ takes effect on the two output qubits. On the other hand, if the input qubit assumes the one state, then $\mathrm{U_1^{(2)} \otimes U_2^{(2)}}$ is applied. This approach ensures that the rotations on the two output qubits are dependent on the states of the input qubits. 
This setup can be constructed according to the circuit depicted in FIG.~\ref{fig:error_circuit}. 

The gates $\mathrm{U}_1^{(i)}$ and $\mathrm{U}_2^{(i)}$ are defined as follows:
$\mathrm{U}_1^{(i)}=\mathrm{H} \, \mathrm{R_z}(-b^i) \, \mathrm{R_z}(-a_d^i x_d) ... \mathrm{R_z}(-a_1^i x_1) \, \mathrm{H} = \mathrm{R_x}(2\phi^i)$ where $\phi^i = (-b^i-a_d^i x_d+...-a_1^i x_1)$ and $\mathrm{U}_2^{(i)}=\mathrm{R_y}(\gamma^i)$. These can be made into conditional gates $\mathrm{CU}^{(i)}_1$ and $\mathrm{CU}^{(i)}_2$ using single qubit rotations and the controlled-z (CZ) gate, operations that the perceptron is well equipped with. The techniques to implement controlled-Z, phase gates, and single-qubit rotations have been previously detailed in Ref.~\cite{AraizaBravo2023}.
The conditional gates $\mathrm{CU}^{(i)}_1$ and $\mathrm{CU}^{(i)}_2$ are constructed as follows:
\begin{equation}
    \mathrm{CU}^{(i)}_1 = \mathrm{CZ}. [\mathrm{I}\otimes\mathrm{R_x}(-\phi^i)].\mathrm{CZ}. [\mathrm{I}\otimes\mathrm{R_x}(\phi^i)]
\end{equation}
\begin{equation}
    \mathrm{CU}^{(i)}_2 = \mathrm{CZ}. [\mathrm{I}\otimes\mathrm{R_y}(-\gamma^i)].\mathrm{CZ}. [\mathrm{I}\otimes\mathrm{R_y}(\gamma^i)]
\end{equation}
In these expressions, the input qubit controls the operations on the output qubits. Up to an arbitrary phase, a QP equipped with single qubit rotations entangles the input and output qubits using a \(\mathrm{CZ}\) gate, which is native to the QP when dealing with input-output qubit pairs, and is used to construct the conditional gates. By establishing this method for constructing the $\mathrm{U}$ gate for a three qubit system, we can extend similar techniques to larger quantum systems. This involves scaling the conditional operations and incorporating more input qubits, maintaining the principles of controlled rotations and entanglement through gates like $\mathrm{CZ}$.

Thus, we have demonstrated how the specified parameterized quantum circuit can be realized using the two output qubit QP model. With this implementation in place, we now focus on the approximation error bounds associated with the function that the circuit encodes.

\section{Approximation Error Bound}\label{sec:approx_error}
As established in Ref. \cite{error}, we consider the application of the variational quantum circuit to learning continuous and integrable functions \( f: \mathbb{R}^d \to \mathbb{R} \). For such functions, the Fourier transform is defined as
\[
\hat{f}(\xi) := \int_{\mathbb{R}^d} e^{-2\pi i y \cdot \xi} f(y) \, dy, \quad \text{for } \xi \in \mathbb{R}^d,
\]
with the \( L_1 \)-norm given by
\[
L_1[\hat{f}] := \int_{\mathbb{R}^d} |\hat{f}(\xi)| \, d\xi,
\]
which can be either finite or infinite. Given the operator \( \mathrm{C_\mathrm{n}}(\theta, \mathbf{x}) \) and a constant \( R > 0 \), the map \( f_{R,n,\theta} : \mathbb{R}^d \to \mathbb{R} \) is defined as in Eq. (\ref{error_map}). To facilitate the analysis, we define the function space
\[
F := \left\{ f : \mathbb{R}^d \to \mathbb{R} \mid f \in C(\mathbb{R}^d) \right\},
\]
and introduce a fixed probability measure \( \mu \) on \( \mathbb{R}^d \) to evaluate the approximation error. It has been shown that the outputs of \( \mathrm{C_n}(\theta, x) \) can approximate functions satisfying \( \hat{f} \in L_1(\mathbb{R}^d) \) up to an error of size \( n^{-\frac{1}{2}} \), highlighting that there is no curse of dimensionality, as the number of qubits required scales logarithmically with \( n \).

In formal terms, according to Theorem 2.4 in Ref.~\cite{error}, for any \( R > 0 \), \( f \in F_R \), and \( n \in \mathbb{N} \), there exists \( \theta \in \Theta \) such that
\[
\left( \int_{\mathbb{R}^d} \left| f(x) - f_{R,n,\theta}(x) \right|^2 \mu(dx) \right)^{1/2} \leq \frac{L_1[\hat{f}]}{\sqrt{n}}.
\]
Since the QP can implement the specified parameterized quantum circuit, it inherits the established lower bound on the approximation error. Applying these error bounds to the QP model establishes its effectiveness for function approximation and offers a framework for assessing its performance. 


\end{document}